\documentclass[reprint,aps,prb,superscriptaddress,notitlepage,nofootinbib,longbibliography,floatfix]{revtex4-2}

\usepackage[T1]{fontenc}
\usepackage{microtype}
\usepackage{bm,bbold}
\usepackage{newtxtext,newtxmath,xspace}
\usepackage{siunitx}
\usepackage{graphicx,xcolor}
\usepackage{tikz}
\usepackage{varwidth}
\usepackage{fancyhdr}
\usepackage{gensymb}
\usepackage[colorlinks=true,
            linkcolor=blue,
            urlcolor=blue,
            citecolor=blue]{hyperref}
\usepackage{soul}
\usepackage{orcidlink}
\setstcolor{red}
\allowdisplaybreaks[2]
\newcommand{\iu}{{\mathtt{i}\mkern1mu}}

\makeatletter
\newenvironment{cellvarwidth}[1][t]
    {\begin{varwidth}[#1]{\linewidth}}
    {\@finalstrut\@arstrutbox\end{varwidth}}
\makeatother

\newcommand{\StateSketch}[3]{%
    \begin{tikzpicture}[x=1bp,y=1bp,line width=0.75bp,line cap=butt,line join=miter]
        \path[use as bounding box] (0,0) rectangle (#1,#2);
        \begin{pgfinterruptboundingbox}
            #3
        \end{pgfinterruptboundingbox}
    \end{tikzpicture}%
}
\newcommand{\StateSketchArrow}[3]{%
    \begin{scope}[shift={#1},rotate=#2]
        \draw (0,0) -- (0,#3);
        \draw[line cap=round] (-3.5,{#3-6}) -- (0,#3) -- (3.5,{#3-6});
    \end{scope}%
}
\newcommand{\SketchFQ}{%
    \StateSketch{33}{19}{%
        \foreach \x/\rx in {4.982/4.582,16.327/4.582,27.891/4.691} {
            \draw (\x,9.727) ellipse [x radius=\rx bp,y radius=2.4bp];
            \draw[dash pattern=on 0.75bp off 0.75bp] (\x,18.236) -- (\x,9.732);
            \draw[dash pattern=on 0.75bp off 0.75bp] (\x,7.327) -- (\x,1.158);
        }
    }%
}
\newcommand{\SketchFP}{%
    \StateSketch{35}{23}{%
        \StateSketchArrow{(6.236,0.119)}{0}{16.265}
        \StateSketchArrow{(17.364,0.338)}{0}{16.265}
        \StateSketchArrow{(28.709,0.338)}{0}{16.265}
    }%
}
\newcommand{\SketchY}{%
    \StateSketch{35}{42}{%
        \draw[gray!70,dash pattern=on 3bp off 2.25bp] (17.455,40.327) -- (17.455,6.127);
        \StateSketchArrow{(17.455,23.146)}{40}{16.265}
        \StateSketchArrow{(17.455,23.146)}{-40}{16.265}
        \StateSketchArrow{(17.455,23.146)}{180}{16.265}
    }%
}
\newcommand{\SketchZAFM}{%
    \StateSketch{30}{37}{%
        \StateSketchArrow{(17.049,15.046)}{-20}{16.265}
        \StateSketchArrow{(17.049,15.077)}{-140}{10.596}
        \StateSketchArrow{(17.453,15.343)}{100}{10.596}
    }%
}
\newcommand{\SketchUmbrella}{%
    \StateSketch{97}{38}{%
        \draw[gray!55,line width=0.3bp] (22.382,22.418) ellipse [x radius=17.5bp,y radius=3.2bp];
        \StateSketchArrow{(22.393,6.198)}{45.456}{22.692}
        \StateSketchArrow{(22.055,5.786)}{-36.651}{17.813}
        \StateSketchArrow{(22.055,6.132)}{-15.892}{20.502}
        \begin{scope}[draw=gray!70]
            \StateSketchArrow{(42.564,18.927)}{-90}{10.879}
        \end{scope}
        \draw[gray!65,line width=0.5bp] (80.25,19.13) circle [radius=15.25bp];
        \foreach \angle in {60,180,300} {
            \begin{scope}[shift={(80.25,19.13)},rotate=\angle,line width=0.55bp]
                \draw[dash pattern=on 0.55bp off 0.55bp] (0,0) -- (14,0);
                \draw[line cap=round] (11.4,-1.4) -- (14,0) -- (11.4,1.4);
            \end{scope}
        }
    }%
}
\newcommand{\SketchV}{%
    \StateSketch{38}{33}{%
        \StateSketchArrow{(15.629,11.922)}{-110}{16.265}
        \StateSketchArrow{(15.785,11.765)}{30}{16.265}
        \StateSketchArrow{(14.032,10.864)}{30}{16.265}
    }%
}
\newcommand{\SketchUUD}{%
    \StateSketch{34}{29}{%
        \StateSketchArrow{(5.673,5.538)}{0}{16.265}
        \StateSketchArrow{(17.018,5.756)}{0}{16.265}
        \StateSketchArrow{(28.364,22.764)}{180}{16.265}
    }%
}

\begin{document}

\title{Thermal and dynamical behaviors of the spin-nematic phases in a spin-1 triangular lattice}

\author{Lei~Xu}
\affiliation{School of Physics, Zhejiang University, Hangzhou 310058, China}

\author{Zhentao~Wang}
\email{ztwang@zju.edu.cn}
\affiliation{Center for Correlated Matter, Zhejiang University, Hangzhou 310058, China}
\affiliation{School of Physics, Zhejiang University, Hangzhou 310058, China}

\begin{abstract}
The triangular lattice compound Na$_2$BaNi(PO$_4$)$_2$ has been recently established as a strong candidate for realizing the spin-nematic physics.
In this theoretical work, we study the finite-temperature properties and the dynamics of a spin-1 triangular lattice model, which reproduces the spin-nematic physics and provides insights relevant to the compound.
We show that a small biquadratic interaction can be used for stabilizing the SN phases in the semiclassical approach with SU(3) coherent states, which allows for a mathematical parameterization of the spin-nematic wave functions.
With a semiclassical Monte Carlo calculation, the finite temperature transitions into the spin-nematic phases are shown to be related to BKT transitions associated with the proliferation of half vortices in the original spin-1 model.
Using quantum Monte Carlo calculations with a low-energy effective model, the phase boundaries are determined and found to agree qualitatively with the semiclassical analysis.
Within the spin wave approach, the low-energy spin dynamics can be captured equally well by either the spin-1 model with a small biquadratic term, or the effective spin-$1/2$ model.
The accidental degeneracy of the semiclassical approach at zero field is revealed by explicit analytical parameterization, and is shown to be associated with the pseudo-Goldstone mode observed in the spin dynamics.
\end{abstract}

\date{\today}
\maketitle

\section{Introduction}
\label{sec:intro}

The search for ``hidden order'' that evades conventional detections has been an ongoing effort in the study of condensed matter physics~\cite{MydoshJA2011_RMP}.
Quantum spin nematic (SN), a macroscopic spin arrangement that spontaneously breaks the spin rotational symmetry while preserving the time-reversal symmetry, is one typical realization of hidden order as it does not couple directly to magnetic probes~\cite{BlumeM1969,MatveevVM1973,AndreevAF1984,PapanicolaouN1988,ChubukovAV1990}.
While initially being a pure theoretical curiosity, the observation of the spin-disordered phase in NiGa$_2$S$_4$~\cite{NakatsujiS2005} sparked the interest and motivated a series of theoretical~\cite{TsunetsuguH2006,LauchliA2006,ShindouR2013_nematic_DSF,SmeraldA2013} and experimental efforts~\cite{SvistovLE2011,OrlovaA2017,NawaK2014,PovarovKY2019,SkoulatosM2019,BhartiyaVK2019,IshikawaH2015,JansonO2016,YoshidaM2017,KohamaY2019,KimH2024} in confirming the existence of the SN phase.

Recently, the triangular lattice (TL) compound Na$_2$BaNi(PO$_4$)$_2$ has emerged as a strong candidate for hosting multiple field-induced SN phases~\cite{LiN2021,DingF2021,ShengJ2025_NiP1,ShengJ2025_NiP2,HuangQ2025}, whose common origin was shown to be from condensation of two-magnon bound states~\cite{ShengJ2025_NiP2}.
A few advantages distinguish this new compound among the candidates: 1.~A low saturation field ($B_\text{s}\sim \qty{1.8}{T}$) allows probes to cover all the low-temperature phases; 2.~Reliable model parameter extraction revealed a 2D nearest-neighbor-only U(1)-symmetric model that describes practically all the spectroscopic signals in all low-temperature phases~\cite{ShengJ2025_NiP1,ShengJ2025_NiP2,HuangQ2025}; 3.~Inelastic neutron scattering (INS) measurements inside the SN phases of this compound revealed low-lying modes that agree with the quadrupole wave excitations~\cite{ShengJ2025_NiP2}.

A few challenges remain in understanding the physics of Na$_2$BaNi(PO$_4$)$_2$.
While the smallness of the magnetic exchange parameters benefits its field tuning, at the same time it reduces the temperature ranges where the SN physics becomes relevant.
For instance, even at $T=\qty{60}{mK}$ only a short-range ordered SN-supersolid phase is detected~\cite{ShengJ2025_NiP1,ShengJ2025_NiP2}, so essentially the phase boundary enclosing the SN-supersolid is still unknown.
While the finite-$T$ boundary between the paramagnetic (PM) phase and the SN phase appears to be clear at high field, the one between the up-up-down (UUD) and the SN phases is almost invisible in the specific heat measurements.
Moreover, the transition between UUD and SN appears to be first order that obscured the INS measurements~\cite{ShengJ2025_NiP2}.

In this paper, we try to understand both the thermal and dynamical behaviors of this compound from a model perspective.
Using semiclassical and quantum Monte Carlo (QMC) calculations at finite temperature, we establish the phase boundaries of Na$_2$BaNi(PO$_4$)$_2$, providing insights into the aforementioned difficulties in understanding its phase diagram.
While conventional spin wave calculations fail to produce meaningful spin excitation spectra for this compound, we show a few approaches to overcome the difficulty, which in turn reveals rich configuration landscape and degeneracy in the zero-field SN-supersolid phase.

%The rest of the paper is organized as follows: In Sec.~\ref{sec:Model}, we introduce the microscopic spin-1 TL model.
%The zero- and finite-temperature phase diagrams of this model are subsequently evaluated utilizing SU(3) coherent states and semi-classical MC simulations in Sec.~\ref{sec:semi}.
%In Sec.~\ref{sec:QMC}, we map the original spin-$1$ model onto an effective spin-1/2 model, and determine the $T>0$ phase diagrams using the stochastic series expansion (SSE) quantum MC (QMC) method.
%In Sec.~\ref{sec:DSSF}, we analyze the dynamic spin structure factors (DSSF) for the ordered phases using semiclassical calculations, and also discuss the degenerate manifold of the zero-field SN-supersolid phase.
%Conclusions and discussions are presented in Sec.~\ref{sec:Discussion}.
%[xx {\color{red} to be updated} xx] Additionally, Appendix~\ref{sec:energy} elaborates on the exact configurations and energies of the semi-classical phases discussed in Sec.~\ref{sec:semi}.
%Appendix~\ref{sec:effective_model} details the derivation procedure of the effective model.

\section{Model}
\label{sec:Model}

For Na$_2$BaNi(PO$_4$)$_2$, the relevant physics was shown to be described by a spin-1 XXZ model with easy-axis single-ion anisotropy on a TL~\cite{ShengJ2025_NiP1}:
\begin{equation}
\label{eq:model}
\mathcal{H}  =J\sum_{\langle ij\rangle}\left(S_{i}^{x}S_{j}^{x}+S_{i}^{y}S_{j}^{y} +\Delta S_{i}^{z}S_{j}^{z}\right) -D\sum_{i}\left(S_{i}^{z}\right)^{2},
\end{equation}
where $J>0$ is a nearest-neighbor antiferromagnetic exchange, $\Delta$ and $D$ parameterize the XXZ exchange anisotropy and single-ion anisotropy, respectively.
%$H$ is the magnitude of the magnetic field applied along the $z$-axis.
In this model, both the $\Delta$ and $D$ terms preserve the continuous U(1) spin-rotational symmetry about the $z$-axis.
%Throughout this paper, we fix the exchange parameters to $J = 1$ and $\Delta = 1$, focusing our investigation on the interplay between $D$, $H$, and thermal fluctuations.
Experimentally, the values were determined as $J=\qty{0.032}{meV}$, $\Delta=1.13$ and $D/J=3.97$~\cite{ShengJ2025_NiP1}.
Huang \textit{et al.} also reported similar fitting results: $J=\qty{0.0348(2)}{meV}$, $\Delta=1.06(1)$, and $D=\qty{0.138(1)}{meV}$ ($D/J\approx 3.97$)~\cite{HuangQ2025}.

This model has been studied by several groups in the past.
In the thesis of T\'oth~\cite{TothTA2011_thesis}, a semiclassical analysis based on SU(3) coherent states was performed on a modified model where biquadratic interactions were also considered:
\begin{equation}
\label{eq:model2}
\mathcal{H}_K = \mathcal{H} + K \sum_{\langle ij\rangle} \left( \bm{S}_i \cdot \bm{S}_j \right)^2.
\end{equation}
Besides establishing a semiclassical phase diagram, Ref.~\cite{TothTA2011_thesis} pointed out that a SN-supersolid phase can be realized by mapping Eq.~\eqref{eq:model} to an effective spin-$\frac{1}{2}$ model in the large-$D/J$ (strongly easy-axis) limit.

Moreno-Cardoner \textit{et al.}~\cite{Moreno-CardonerM2014} further clarified that, at $\Delta=1$ and with large $D/J$, two different SN-supersolid phases exist for opposite signs of $K$.
In addition, they found a region between the two SN-supersolids where cluster mean-field (CMF) calculations become unstable, possibly related to the formation of quantum spin liquids proposed in an earlier work~\cite{SerbynM2011}.
In fact, this region is centered around the $K=0$ line, i.e., CMF is unstable for the original model Eq.~\eqref{eq:model} for a wide parameter region when $D$ is finite.
In particular, for $\Delta \sim 1$, $D/J\sim 4$ relevant to the TL material Na$_2$BaNi(PO$_4$)$_2$~\cite{ShengJ2025_NiP1,ShengJ2025_NiP2,HuangQ2025}, the phase diagram in Ref.~\cite{Moreno-CardonerM2014} puts a question mark indicating this numerical instability of the CMF method.

More recently, Seifert and Savary~\cite{SeifertUFP2022} performed an extensive study of the phase diagram of model \eqref{eq:model2} using SU(3) coherent states within a three-sublattice ansatz.
In particular, the multidimensional parameter space spanned by \{$\Delta$, $D/J$, $K/J$\} was studied systematically, revealing phase diagrams hosting a rich family of semiclassical states.
Importantly, they discovered that a semiclassical state (named $z$AFM$_\text{deg}$ therein) has an accidental U(1) symmetry, and argued that it becomes a SN-supersolid state once quantum fluctuations are considered.

Those early theoretical studies were performed before the discovery of SN phases in Na$_2$BaNi(PO$_4$)$_2$.
As such, they put more emphasis on exploring the zero-temperature and zero-field phases by playing with the anisotropy, competing bilinear and biquadratic exchange parameters.
In this paper, we dive deeper into the parameter region relevant to Na$_2$BaNi(PO$_4$)$_2$ on several fronts.
In particular, we consider a magnetic field applied along the $c$-axis:
\begin{equation}\label{eq:H_tot}
\mathcal{H}_\text{tot} \equiv \mathcal{H} + K \sum_{\langle ij\rangle} \left( \bm{S}_i \cdot \bm{S}_j \right)^2 - H \sum_i S_i^z,
\end{equation}
where $H$ is the Zeeman coupling coefficient.
The U(1) symmetry is intact in $\mathcal{H}_\text{tot}$ which is essential for the 2-magnon condensations reported in Refs.~\cite{ShengJ2025_NiP1,ShengJ2025_NiP2}.
The biquadratic coefficient $K$ will be set to either a small number or zero, so the dominant physics of Na$_2$BaNi(PO$_4$)$_2$ is retained.

%{\color{orange}
%[xx defer after model xx]
%While finite-temperature phenomena---such as Berezinskii-Kosterlitz-Thouless (BKT) transitions and multiple intermediate thermodynamic phases---have been investigated in classical and semi-classical limits of related isotropic or simple XXZ triangular models,~\cite{SeabraL2011,GvozdikovaMV2011}
%}

%{\color{orange}
%To unambiguously identify and distinguish the SN and SN-supersolid phases, we calculate the out-of-plane dipolar spin structure factor $\mathcal{S}_{\text{d}}^{zz}\left(\bm{k}\right)$ and the transverse quadrupolar structure factor $\mathcal{S}_{\text{q}}^{\perp}\left(\bm{k}\right)$, defined as
%\begin{subequations}
%\begin{align}
%\mathcal{S}_{\text{d}}^{zz}\left(\bm{k}\right)& =\frac{1}{N}\sum_{i,j}e^{-\iu\bm{k}\cdot\left(\bm{r}_{i}-\bm{r}_{j}\right)}\langle S_{i}^{z}S_{j}^{z}\rangle, \\
%\mathcal{S}_{\text{q}}^{\perp}\left(\bm{k}\right) & = \frac{1}{N}\sum_{i,j} e^{-\iu\bm{k}\cdot\left(\bm{r}_{i}-\bm{r}_{j}\right)} \langle {Q}_{i}^{x^{2} - y^{2}} {Q}_{j}^{x^{2} - y^{2}} + Q_{i}^{xy} Q_{j}^{xy} \rangle,
%\end{align}
%\end{subequations}
%where $N = L \times L$ is the total number of lattice sites.
%The relevant transverse quadrupolar components in the Cartesian basis are concisely denoted as $Q_{i}^{x^{2} - y^{2}} \equiv (S_{i}^{x})^2-(S_{i}^{y})^2$ and $Q_{i}^{xy} \equiv \{ S_{i}^{x},S_{i}^{y} \}$.
%}

For convenience, we shall define the charge $m$ of an operator under the conserved generator $\sum_i S_i^z$, where
\begin{equation}
\left[ S^z, O^m \right] = m O^m.
\end{equation}
Under this classification, we have three groups of operators: $m=\pm 1$ for $S^{\pm}$ and $Q^{\pm 1}$; $m=\pm 2$ for $Q^{\pm 2}$; and $m=0$ for $S^z$ and $Q^{3z^2-r^2}$.
Note that the quadrupolar operators are defined as
\begin{subequations}
\begin{align}
Q^{\pm 2} &\equiv Q^{x^2-y^2} \pm \iu Q^{xy} = \left(S^{\pm}\right)^2, \\
Q^{\pm 1} &\equiv Q^{xz} \pm \iu Q^{yz} = S^{\pm}S^z + S^z S^{\pm},\\
Q^{3z^2-r^2} &\equiv \left[ 3\left(S^z\right)^2 - S (S+1) \right] / \sqrt{3}.
\end{align}
\end{subequations}

\section{Semi-classical analysis}
\label{sec:semi}

We start by employing the standard semi-classical variational method based on SU(3) coherent states to study the $T=0$ phase diagram of Eq.~\eqref{eq:H_tot}, essentially the same approach as in Ref.~\cite{SeifertUFP2022} except that we relax the three-sublattice ($\sqrt{3}\times \sqrt{3}$) ansatz to magnetic unit cells of size $3\times 3$ and $6 \times 6$.
For all results shown in this paper, the three types of magnetic unit cells yield identical results.
%The larger variational space in larger magnetic unit cell will provide further assurance of the results obtained from the 3-sublattice case, and may find additional lower-energy solutions in some of the parameter region.

\begin{figure}
\centering
\includegraphics[width=\columnwidth]{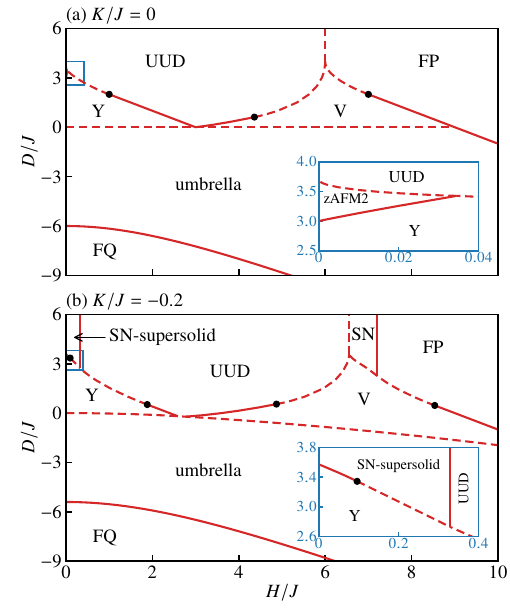}
\caption{Semiclassical phase diagram of the model Eq.~\eqref{eq:H_tot} at $T=0$, where we set the exchange anisotropy $\Delta=1$.
(a) $K/J=0$. (b) $K/J=-0.2$.
Solid (dashed) red lines indicate continuous (first order) phase boundaries, and the dots are tricritical points.
The insets show zoomed-in views of the narrow regions at low field.}
\label{fig:phd_T0}
\end{figure}

\begin{table*}[tbp]
\centering
\caption{States identified in Fig.~\ref{fig:phd_T0} (except for SN-supersolid and SN which are discussed explicitly in the main text). The subscript $s\in\{ A,B,C\}$ distinguishes the three sublattices, and $R^{\alpha}(\phi)$ denotes an SO(3) rotation matrix about the $\alpha$-axis by an angle $\phi$.}
\label{tab:state}
\vskip 1.5mm
\begin{tabular}{cccc}
\hline
Sketch & Name & \parbox{4.6cm}{Description} & \parbox{6.5cm}{Representative configuration}\tabularnewline
\hline
\hline
\begin{cellvarwidth}[t]
    \centering
    \

    \SketchFQ

    \
\end{cellvarwidth} & FQ & \parbox{4.6cm}{Ferroquadrupolar state.} & \parbox{6.5cm}{$\bm{d}_{s}=\left(0,0,1\right)^{\text{T}}$.}

\tabularnewline

\begin{cellvarwidth}[t]
    \centering
    \

    \SketchFP

    \
\end{cellvarwidth} & FP & \parbox{4.6cm}{Fully polarized state.} & \parbox{6.5cm}{$\bm{d}_{s}=\left(-\iu,1,0\right)^{\text{T}}/\sqrt{2}$.}

\tabularnewline

\begin{cellvarwidth}[t]
    \centering
    \

    \SketchY

    \
\end{cellvarwidth} & Y & \parbox{4.6cm}{Coplanar `Y'-shaped state.} & \parbox{6.5cm}{$\bm{d}_{s}=R^{x}\left(\phi_{s}\right)\left(-\iu\cos\eta_{s},\sin\eta_{s},0\right)^{\text{T}}$, with $\phi_{A}=-\phi_{B},\eta_{A}=\eta_{B}$ and $\phi_{C}=0$.}
\tabularnewline

\begin{cellvarwidth}[t]
    \centering
    \

    \SketchZAFM

    \
\end{cellvarwidth} & zAFM2 & \parbox{4.6cm}{Coplanar state similar to `Y', but lacking any moment strictly parallel to the $z$-axis~\cite{SeifertUFP2022}.} & \parbox{6.5cm}{$\bm{d}_{s}=R^{x}\left(\phi_{s}\right)\left(-\iu\cos\eta_{s},\sin\eta_{s},0\right)^{\text{T}}$, with all parameters unconstrained.}
\tabularnewline
\begin{cellvarwidth}[t]
    \centering
    \

    \SketchUmbrella

    \
\end{cellvarwidth} & umbrella & \parbox{4.6cm}{Umbrella state where all moments maintain an identical canting angle with respect to the $z$-axis, while their in-plane components form a $120^{\circ}$ configuration.} & \parbox{6.5cm}{$\bm{d}_{s}=R^{z}\left(\theta_{s}\right)R^{x}\left(\phi_{s}\right)\left(-\iu\cos\eta_{s},\sin\eta_{s},0\right)^{\text{T}}$, with $\phi_{A}=\phi_{B}=\phi_{C},\eta_{A}=\eta_{B}=\eta_{C}$ and $\theta_{s}=0, 2\pi/3, 4\pi/3.$}
\tabularnewline

\begin{cellvarwidth}[t]
    \centering
    \

    \SketchV

    \
\end{cellvarwidth} & V & \parbox{4.6cm}{Coplanar `V'-shaped state.} & \parbox{6.5cm}{$\bm{d}_{s}=R^{x}\left(\phi_{s}\right)\left(-\iu\cos\eta_{s},\sin\eta_{s},0\right)^{\text{T}}$, with $\phi_{A}=\phi_{B}$ and $\eta_{A}=\eta_{B}$.}
\tabularnewline

\begin{cellvarwidth}[t]
    \centering
    \

    \SketchUUD

    \
    \end{cellvarwidth} &
    \begin{cellvarwidth}[t]
        \centering
        UUD
\end{cellvarwidth} & \parbox{4.6cm}{Collinear up-up-down state.} & \parbox{6.5cm}{$\bm{d}_{s}=R^{x}\left(\phi_{s}\right)\left(-\iu,1,0\right)^{\text{T}}/\sqrt{2}$, with $\phi_{A}=\phi_{B}=0$ and $\phi_{C}=\pi$.}
\tabularnewline
\hline
\end{tabular}
\end{table*}

Within this approach, the total wave function is approximated by a site-factorized product state
$| \Psi \rangle = \prod_{i}| \Psi_i \rangle$,
where the local state $|\Psi_i \rangle$ is parameterized by a 3-vector in the $\mathbb{CP}^2$ space:
\begin{equation}
|\Psi_{i} \rangle = \{ |1\rangle,\,  |0\rangle,\, |-1\rangle \} \bm{Z}_i
= \{ |x\rangle,\,  |y\rangle,\, |z\rangle \} \bm{d}_i,
\end{equation}
where the two bases are related by a unitary transformation.
While the variational space is equally described by the dipolar ($\bm{Z}_i$) and the quadrupolar ($\bm{d}_i$) bases, we adopt the quadrupolar language in this Article for its convenience in describing the quadrupolar physics, unless specified otherwise.
The overall phase of $\bm{d}_i$ is fixed by $\text{Im}(\bm{d}_i \cdot \bm{d}_i) = 0$ in this Article (see Appendix~\ref{sec:conversion}).

The $T=0$ phase diagram at zero field has been established in Seifert and Savary~\cite{SeifertUFP2022}.
When the biquadratic interaction is negligible, five semiclassical phases are present at the Heisenberg point ($\Delta=1$) when the single-ion anisotropy $D$ is tuned.

Figure~\ref{fig:phd_T0} shows the evolution of the phases in the presence of a finite magnetic field, where the two panels show results at $K/J=0$ and $K/J=-0.2$, respectively.
A rich set of phases can be identified (see Table~\ref{tab:state} except for SN-supersolid and SN which will be elaborated explicitly in the following text), and different types of transitions (continuous, first order) can be seen.
Interestingly, several tricritical points, as originally proposed in He$^3$ - He$^4$ mixtures~\cite{LandauLD1937,GriffithsRB1970}, appear in both phase diagrams.

While SN-supersolid and SN phases are expected to be present~\cite{ShengJ2025_NiP1} at $K=0$ in an accurate quantum treatment, they do not appear in any region in Fig.~\ref{fig:phd_T0}(a), already shown to be an artifact from the semiclassical SU(3) approximations~\cite{ShengJ2025_NiP2}.
Figure~\ref{fig:phd_T0}(b) shows that a small biquadratic interaction is able to enhance quadrupolar orders and produce finite regions of SN phases with easy-axis single-ion anisotropy ($D>0$).
We emphasize that, in the experiments for Na$_2$BaNi(PO$_4$)$_2$ there is no evidence of a finite $K$ term~\cite{ShengJ2025_NiP1,ShengJ2025_NiP2},
and the purpose of keeping it in this paper is mainly for building intuitive physical pictures within the semiclassical SU(3) approach.

More specifically, the energy minimization procedure demonstrates that both SN phases relevant to Na$_2$BaNi(PO$_4$)$_2$ can be characterized by the directors:
\begin{equation}\label{eq:director-SN}
\bm{d}_s =
\begin{pmatrix}
\cos \chi_s \cos \phi_s +\iu \sin \chi_s \sin \phi_s \\
\cos \chi_s \sin \phi_s -\iu \sin \chi_s \cos \phi_s \\
0
\end{pmatrix},
\end{equation}
where $s=\text{A},\,\text{B},\,\text{C}$ for the 3 sublattices.
For the low-field SN-supersolid phase, we have $\bm{d}_B=\bm{d}_C\neq \bm{d}_A$, and for the high-field SN phase we have $ \bm{d}_A =  \bm{d}_B =  \bm{d}_C$.

Accordingly, the dipole moments are evaluated to
\begin{equation}
\langle \bm{S}_s \rangle =
(0, 0, -\sin 2\chi_s),
\end{equation}
and quadrupole moments are
\begin{equation}
\langle Q_s^{\pm 2} \rangle = -\cos 2\chi_s e^{\pm 2\iu \phi_s},\quad \langle Q_s^{\pm 1} \rangle = 0.
\end{equation}

At this point, the physical meaning of the parameterization is rather clear: $\phi_s$ represents the spontaneous broken U(1) in-plane angle, and $\chi_s$ controls the magnitudes of the in-plane quadrupole and out-of-plane dipole moments.

We note that at $H=0$, there is an accidental degeneracy for $D/J \gg 1$, already reported by Seifert and Savary~\cite{SeifertUFP2022}, where they named the family of degenerate states as zAFM$_{\mathrm{deg}}$, and argued that supersolidity can be realized in the presence of an infinitesimal magnetic field.
The analytical structure of the degeneracy was not given in Seifert and Savary's paper, though.

Following the trick by Kleine \textit{et al.}~\cite{KleineB1992} which analyzed the degeneracy of the SU(2) semiclassical states of the antiferromagnetic XXZ model on TL~\cite{MiyashitaS1985}, we found that the energy of the degenerate state at $H=0$ can be written as
\begin{equation}
\begin{split}
e &= A \left[ \frac{1+\lambda}{4} \left(m_x^2 + m_y^2 - 3 \right) + \frac{1-\lambda}{2}m_y \right]\\
&\quad + \frac{9}{2}K-D,
\end{split}
\end{equation}
where
\begin{equation}
A \equiv \Delta J - K/2,\quad
B \equiv K/2,\quad
\lambda \equiv B/A,
\end{equation}
essentially Kleine~\cite{KleineB1992} and Murthy's~\cite{MurthyG1997} results generalized to the $S=1$ model.
The accidental degeneracy is clear by counting the number of degrees of freedom: the degenerate state can be parameterized by three independent variables [after fixing $\phi_s$ in Eq.~\eqref{eq:director-SN} to remove the real Goldstone mode], while the energy of the degenerate state only depends on two independent variables ($m_x$ and $m_y$). Therefore, the energy minimum is generically a one-parameter manifold that parameterizes the accidental degeneracy (see Appendix~\ref{sec:degeneracy}).

\begin{figure}
\centering
\includegraphics[width=\columnwidth]{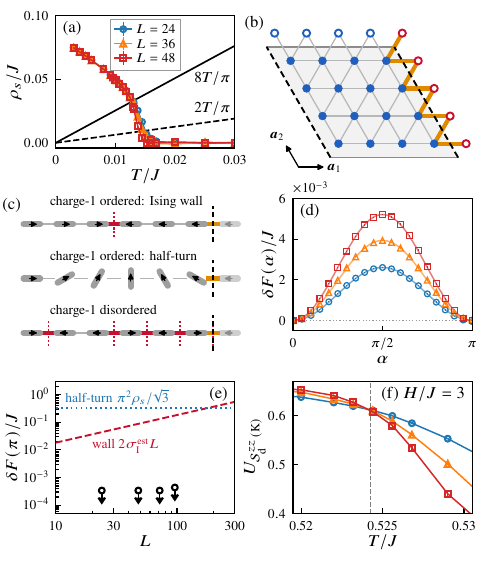}
\caption{Semiclassical Monte Carlo results of the model Eq.~\eqref{eq:H_tot}, with parameters $\Delta=1$, $D/J=5$, and $K/J=-0.2$. (a) Spin stiffness at $H/J=6.75$. The solid and dashed lines indicate the potential Nelson--Kosterlitz jump~\cite{NelsonDR1977} associated with half vortices and full vortices, respectively.
(b) Schematic plot of the cut (dashed lines) and seam (orange solid bonds) on a torus, where the bonds in the seam are twisted in the charge-1 channel as described in the text.
(c) Schematic plot of three possibilities in the presence of a $\pi$-twist at the seam (orange bond): charge-1 sector being ordered with an Ising domain wall (red bond) or through a half turn, or the charge-1 sector being disordered. The gray bars are the directors, and the arrows indicate the charge-1 phase ($s=\pm 1$) of $e^{\iu \theta} = s e^{\iu \phi}$ where $\phi \in [0,\pi)$.
(d) Free energy cost as a function of the seam angle $\alpha$ defined in Eq.~\eqref{eq:twist_seam} for $L=\{24,36,48\}$ at $H/J=6.75$ and $T/J=0.008$.
The solid lines are $\sin^2 \alpha$ scaled to each maximum.
(e) Estimated upper limits of $\delta F(\pi)$ (empty circles with arrowheads) compared to the costs of two ordered scenarios at $H/J=6.75$ and $T/J=0.008$.
(f) Binder parameter of $\mathcal{S}^{zz}_\text{d}(\mathrm{K})$ at $H/J=3$. The vertical dashed line marks the critical temperature $0.5243(7)$, obtained from the finite-size collapse of the Binder parameter with $\nu=5/6$ (see Appendix~\ref{sec:boundaries}).
}
\label{fig:stiffness}
\end{figure}

We now turn to the finite-$T$ analysis of the spin-1 model.
For the antiferromagnetic Heisenberg model with classical spins, the finite-$T$ phase diagram on the TL has been extensively studied by classical MC simulations~\cite{SeabraL2011, GvozdikovaMV2011}.
While quantum fluctuations are completely dropped in these calculations, they still provide valuable insights into the $S=1/2$ spin models on the TL, as the dominant physics often happens in the spin dipole channel there.
For example, the classical MC calculations were employed in Ref.~\cite{GaoY2022_NBCP} to understand finite-T behaviors of the spin-1/2 compound Na$_2$BaCo(PO$_4$)$_2$.

For our $S=1$ model, we performed finite-$T$ semiclassical MC simulations based on SU(3) coherent states and Metropolis updates~\cite{StoudenmireEM2009,RemundK2022},
where the update step was tuned during thermalization to an acceptance rate of about $40\%$ and then fixed.
In these simulations, we used $L \times L$ TLs with periodic boundary conditions (PBC) along both directions.
Each simulation is a temperature scan (cooled from a random state or warmed from the $T=0$ state),
with at least $2\times 10^{4}$ thermalization sweeps and $10^{5}$ measurement sweeps.

In two dimensions, the onset of the quasi long-range~\cite{MerminND1966} quadrupole orders can be captured by the Nelson--Kosterlitz jump~\cite{NelsonDR1977} of the spin stiffness~\cite{FisherME1973}:
\begin{equation}
\rho_{s} = \frac{1}{\mathcal{A}} \left. \frac{\partial^2 F\left(\varphi \right)}{\partial \varphi^{2}} \right|_{\varphi=0},
\end{equation}
where $\mathcal{A}=\sqrt{3}N/2$ is the total area of the TL, $\varphi$ is the twist angle, and each state is twisted along the $z$-axis via:
\begin{equation}
|\Psi_i\rangle \rightarrow e^{-\iu \theta_i S_i^z} |\Psi_i\rangle,\quad \theta_i \equiv \varphi (\hat{\bm{e}} \cdot \bm{r}_i),
\end{equation}
where $\hat{\bm{e}}$ is an arbitrary unit vector.
See Appendix~\ref{sec:stiffness} for the explicit form of $\rho_s$.

Figure~\ref{fig:stiffness}(a) shows the temperature dependence of $\rho_s$ at $\Delta=1$, $D/J=5$, $K/J=-0.2$, and $H/J=6.75$, where the $T=0$ ground state is SN.
Indeed, a jump can be seen at finite $T$.
However, as we only have access to finite system size, it is not  clear whether the jump occurs at $8T_c/\pi$ associated with proliferation of half vortices (quadrupole physics)~\cite{NelsonDR1977,LeeDH1985,CarpenterDB1989,MukerjeeS2006,HuangCJ2025}, or $2T_c/\pi$ for full vortices (dipole physics)~\cite{NelsonDR1977,TeitelS1983,WeberH1988}.

To address this question, we computed the free energy cost of a seam, i.e., the $2L$ bonds crossing a non-contractible cut of the torus [orange bonds and dashed line in Fig.~\ref{fig:stiffness}(b)].
Since $\mathcal{H}_\text{tot}$ conserves $\sum_i S_i^z$, each bond term splits into channels of definite charge,
\begin{equation}
h_{ij} = h_{ij}^{(0)} + \sum_{m=1,2}\left[ h_{ij}^{(m)} + \text{h.c.} \right],
\end{equation}
where $h_{ij}^{(m)}$ collects the products of a charge-$(+m)$ operator on site $i$ and a charge-$(-m)$ operator on site $j$.
A seam of angle $\alpha$ twists only the charge-1 channel of the seam bonds,
\begin{equation}\label{eq:twist_seam}
h_{ij}^{(1)} + \text{h.c.} \rightarrow e^{\iu\alpha} h_{ij}^{(1)} + \text{h.c.}, \quad (ij)\in\text{seam},
\end{equation}
and leaves every other term unchanged.

$\alpha=\pi$ reverses the sign of the charge-1 coupling across the cut.
%, while the quadrupolar (charge-2) channel, for which a $\pi$ rotation gives $e^{2\iu\pi}=1$, is unaffected.
If the charge-1 (dipolar) sector were quasi-long-range ordered, the $\pi$ seam would cost at least the smaller of two contributions [Fig.~\ref{fig:stiffness}(c)]: an Ising domain wall wrapping the torus, $2\sigma_\text{I}L$ ($\sigma_\text{I}$ is the Ising wall tension per bond), or a half-turn of the director spread over the system, $\pi^2\rho_s/\sqrt{3}$.
%Half vortices would then be confined, and the jump would occur at $2T_c/\pi$.
If instead the charge-1 sector is disordered, then the half vortices are free, and the jump occurs at $8T_c/\pi$ instead of $2T_c/\pi$.

The free energy cost is obtained by thermodynamic integration,
\begin{equation}
\delta F(\alpha) \equiv F(\alpha)-F(0) = \int_0^{\alpha} \left\langle \frac{\partial E}{\partial \alpha'} \right\rangle_T \mathrm{d}\alpha' .
\end{equation}
Figure~\ref{fig:stiffness}(d) shows the value of $\delta F(\alpha)$ at $T/J=0.008$, for $L=\{24,36,48\}$.
$\delta F(\pi)$ is practically zero and there is no obvious size dependence.
To see it more clearly, in Fig.~\ref{fig:stiffness}(e) we show the estimated upper bounds of free energy cost up to $L=96$.
The upper bounds of $\delta F(\pi)$ for all computed $L$ are below the estimated cost of an Ising domain wall and the half-turn shown in Fig.~\ref{fig:stiffness}(c), suggesting that the dipolar sector is indeed disordered and that $8T_c/\pi$ should be used to pin down the BKT transition associated with half vortices.
Further details of the $\delta F(\alpha)$ calculation can be found in Appendix~\ref{sec:twist}.

\begin{figure}
\centering
\includegraphics[width=\columnwidth]{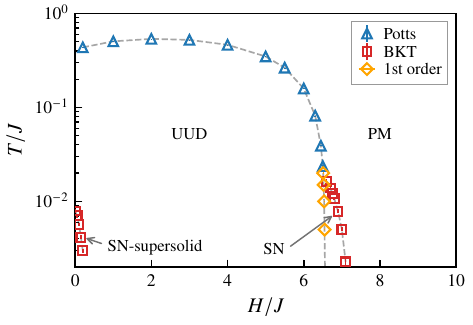}
\caption{Semiclassical phase diagram of the model Eq.~\eqref{eq:H_tot} obtained from Monte Carlo simulation of SU(3) coherent states, with $\Delta=1$, $D/J=5$, and $K/J=-0.2$.
The dashed lines are guides to the eye.}
\label{fig:phd_SU3}
\end{figure}

Both the UUD state and the SN-supersolid state break the lattice translational symmetry in the $S^z$ sector, characterized by the dipole structure factor at the $\bm{k}=\text{K}$ point:
\begin{equation}
\mathcal{S}^{zz}_\text{d}(\bm{k}) \equiv \frac{1}{N}\sum_{i,j}e^{-\iu\bm{k}\cdot\left(\bm{r}_{i}-\bm{r}_{j}\right)}\langle S_{i}^{z}S_{j}^{z}\rangle .
\end{equation}

The corresponding Binder cumulant is defined as~\cite{BinderK1981_RG1,BinderK1981_RG2}:
\begin{equation}\label{eq:Binder}
U_{\mathcal{S}_\text{d}^{zz}(\text{K})} = 1 -\frac{\langle {\mathcal{S}_\text{d}^{zz} (\text{K})}^{2}\rangle}{3\langle \mathcal{S}_\text{d}^{zz} (\text{K}) \rangle^{2}}.
\end{equation}
As shown in Fig.~\ref{fig:stiffness}(f), the crossing of the Binder cumulants at different $L$ can be used to estimate the transition temperature for the $S^z$ sector.

Figure~\ref{fig:phd_SU3} summarizes the semiclassical finite-$T$ phase diagram at $\Delta=1$, $D/J=5$, and $K/J=-0.2$.
The higher temperature boundary with three-state Potts universality class was obtained by the finite-size collapse of the Binder parameter \eqref{eq:Binder}, while the lower temperature BKT boundaries associated with the quadrupolar sector were determined by logarithmic extrapolation of the stiffness crossing temperatures at $8T/\pi$.
In addition, the first order boundary between the UUD state and the SN state was identified by observing the evolution of the prepared half-UUD half-SN state in the semiclassical MC simulations.
Further details can be found in Appendix~\ref{sec:boundaries}.

We note that at $T=0$ the Hamiltonian is invariant under the spin flip $S^z\rightarrow -S^z$.
Combined with the $Z_3$ lattice translations, the $S^z$ order parameter has a $Z_6$ symmetry, so the three-state Potts scaling used at $H>0$ no longer applies.
Instead, an intermediate critical phase bounded by two BKT transitions is expected~\cite{JoseJV1977,IsakovSV2003}.

%\textcolor{red}{[xx update below xx]}
%$T_{c1}$ is strongly renormalized in the finite-size calculations, and we correct it use the finite-size scaling by logarithmic corrections $T_{c1}(L) = T_{c1} [1 + (\log L/L_{0})^{-2}]$.~\cite{WeberH1988, LaflorencieN2007}
%As shown in Figsxxx(c) and (d), by fitting the parameters $L_0$ and $T_{c1}$, we obtain $T_{c1} = 0.178$ at $H=0$.

\section{Quantum Monte Carlo}
\label{sec:QMC}

While the semiclassical SU(3) approach facilitates qualitative understanding of these phases, it is somewhat artificial to include an additional $K\neq 0$ term to allow the SN phases to occupy finite regions in the phase diagram.

In this section, we stick to the strict $K=0$ case.
As demonstrated in Refs.~\cite{ShengJ2025_NiP1,ShengJ2025_NiP2}, the low-energy physics of Na$_2$BaNi(PO$_4$)$_2$ is dominated by the $S_i^z=\pm 1$ doublets for strong easy-axis single-ion anisotropy, allowing for an effective $s=1/2$ description of the low-energy physics:
\begin{equation}
\mathcal{H}_{\text{eff}}=\tilde{J}\sum_{\langle ij\rangle}\left(s_{i}^{x}s_{j}^{x}+s_{i}^{y}s_{j}^{y}+\tilde{\Delta}s_{i}^{z}s_{j}^{z}\right)-\tilde{H}\sum_{i}s_{i}^{z},
\label{eq:Heff}
\end{equation}
where
\begin{equation}
\tilde{J}=-J^{2}/D,\, \tilde{\Delta} =- \left(1+4\Delta D/J \right),\, \tilde{H}=2H.
\label{eq:eff_param}
\end{equation}
See Appendix~\ref{sec:effective_model} for the derivation.

Since the off-diagonal part of the exchange is negative ($\tilde{J}<0$), the effective $s=1/2$ model is sign-problem free for the QMC techniques such as the stochastic series expansion (SSE) method~\cite{SandvikAW1991,SandvikAW1992,SandvikAW1999_SSE,SyljuasenOF2002}.
Consequently, the finite-$T$ phase diagram of the original $S=1$ model can be inferred conveniently from the effective $s=1/2$ model \eqref{eq:Heff}.
In fact, the $T=0$ phases of the $s=1/2$ model have been studied extensively via QMC methods, where the existence of spin supersolidity was reported~\cite{BoninsegniM2005,HeidarianD2005,MelkoRG2005,WesselS2005}.

We performed SSE calculations with global directed-loop updates~\cite{SandvikAW1991,SandvikAW1992,SandvikAW1999_SSE,SyljuasenOF2002} for the effective $s=1/2$ model, on $L\times L$ TLs with $L = \{12,18,24\}$ and PBC along both directions.
In our simulation, each MC sweep consists of one directed-loop update and one diagonal update.
We typically performed $5\times10^{6}$ sweeps, with the first half for thermalization and the rest for measurement.
At each parameter set we performed 10 independent runs for estimating the statistical error.

\begin{figure}
\centering
\includegraphics[width=\columnwidth]{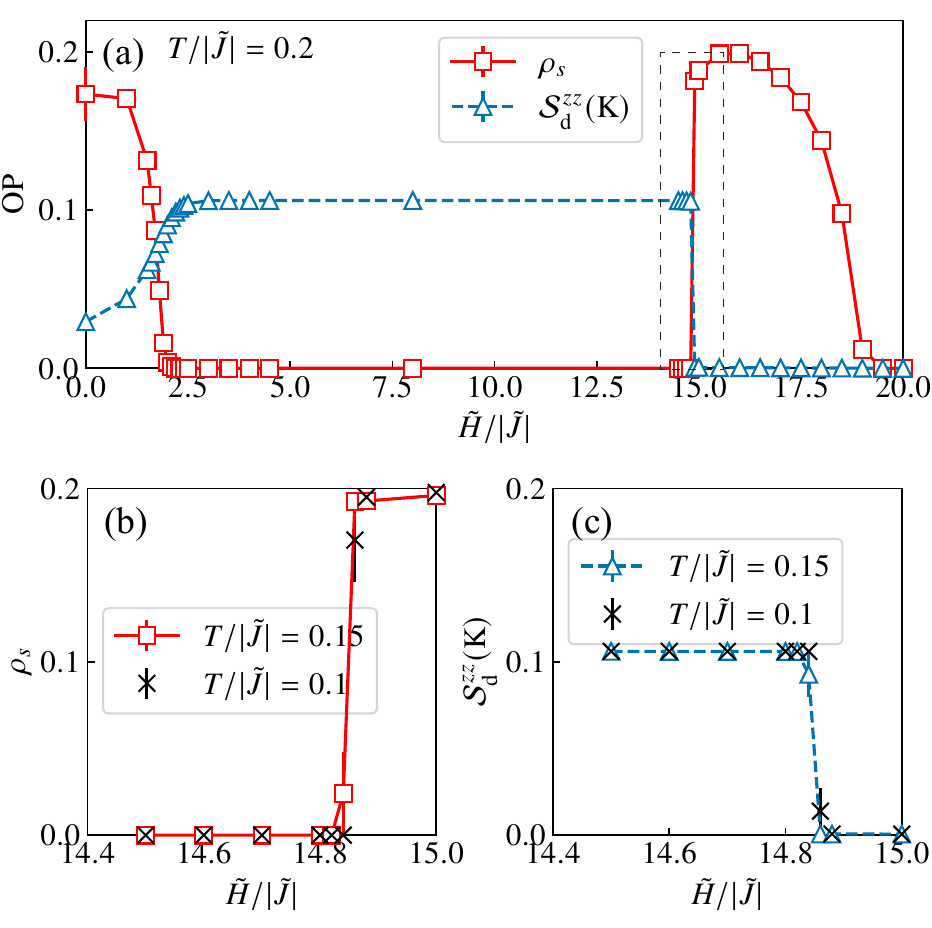}
\caption{SSE results of the effective $s=1/2$ model at $\tilde{\Delta}=-5.5$ and $L=24$.
(a) Field-dependent stiffness $\rho_{s}$ and structure factor $\mathcal{S}_{\text{d}}^{zz}({\text{K}})$ at $T/|\tilde{J}|=0.2$.
(b)(c) Field-dependent stiffness $\rho_{s}$ and structure factor $\mathcal{S}_{\text{d}}^{zz}({\text{K}})$ at lower temperatures near the UUD/SN transition [see dashed rectangle in (a)].}
\label{fig:SSE_A}
\end{figure}

Figure~\ref{fig:SSE_A} shows a typical plot of the order parameters at finite temperature via the QMC of SSE type.
Overall, we see that the results are consistent with the original spin-1 model, showing a sequence of phases including SN-supersolid, UUD, SN, and FP as the field is increased.
Note the jump of both the stiffness $\rho_s$ and the structure factor $\mathcal{S}_{\text{d}}^{zz}({\text{K}})$ as highlighted by the dashed rectangle [Fig.~\ref{fig:SSE_A}(a)], indicating the UUD-SN transition to be first order.
Moreover, in Figs.~\ref{fig:SSE_A}(b)(c) we see that the critical field of this transition remains practically at the same place for multiple temperatures.

To obtain the critical temperatures, we followed similar procedures as in the previous section, i.e., analyzing the finite-size behavior of $\rho_{s}$~\cite{PollockEL1987,SyljuasenOF2002} and $\mathcal{S}^{zz}_\text{d}(\text{K})$ that characterize the two types of transitions.
However, we should note that at the effective model level, the BKT transition is associated with the full vortex in the $s=1/2$ language, so $2T_c/\pi$ should be used here.

\begin{figure}
\centering
\includegraphics[width=\columnwidth]{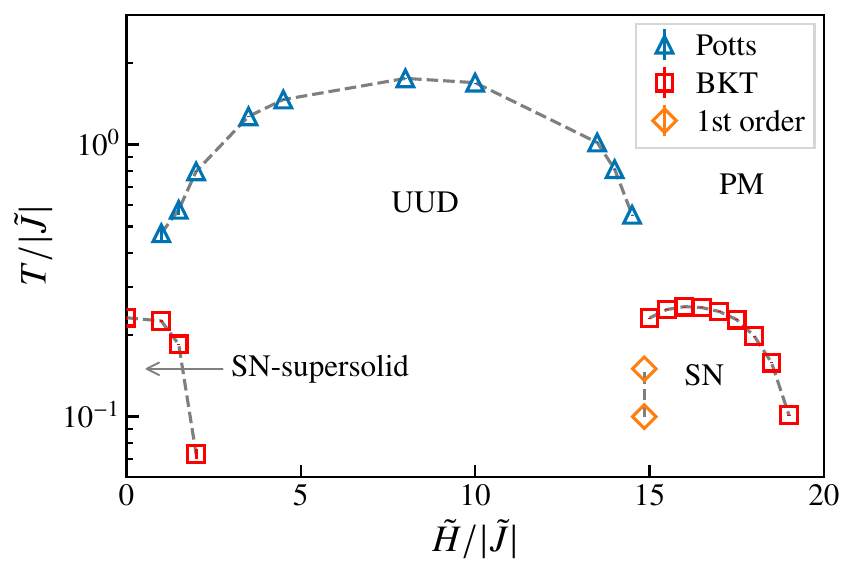}
\caption{QMC phase diagrams of the effective $s=1/2$ model Eq.~\eqref{eq:Heff} at $\tilde{\Delta} = -5.5$.
The red filled circles indicate the critical fields at $T=0$.
The gray dashed lines are guides to the eye.}
\label{fig:phd_SSE}
\end{figure}

Figure~\ref{fig:phd_SSE} shows the finite-temperature phase diagrams of the effective $s=1/2$ model.
%Firstly, at $\tilde{H}=0$ we see that the SN-supersolid phase only appears for relatively strong anisotropies $|\tilde{\Delta} | \gtrsim 4.4$ [Fig.~\ref{fig:phd_SSE}(b)].
%However, the BKT transition temperature also decreases for larger $|\tilde{\Delta}|$ which challenges the QMC convergence.
Note that instead of using realistic parameters ($\tilde{\Delta} \sim -19$) for the Na$_2$BaNi(PO$_4$)$_2$ compound, due to high numerical cost at large $|\tilde{\Delta}|$, here we set $\tilde{\Delta} = -5.5$ where SN-supersolid survives at relatively high temperature to capture the key physics at the qualitative level.
Overall, Fig.~\ref{fig:phd_SSE} agrees qualitatively with the semiclassical finite-$T$ phase diagram Fig.~\ref{fig:phd_SU3}.
And both theoretical phase diagrams agree with the low-temperature experimental results of the Na$_2$BaNi(PO$_4$)$_2$ compound, showing the sequence of the SN-supersolid $\rightarrow$ UUD $\rightarrow$ SN $\rightarrow$ FP at low temperatures.
The boundary between SN-supersolid and UUD phases has not been accurately identified in the Na$_2$BaNi(PO$_4$)$_2$ experiments yet, and our work now provides a qualitative prediction on the shape of that boundary.
In addition, both MC calculations reveal that the UUD/SN boundary is an almost vertical first order line.
In other words, observation of specific heat anomaly associated with the UUD/SN boundary would require extremely fine magnetic field steps in the experiment.

\section{Spin Dynamics}
\label{sec:DSSF}

The spin dynamics of Na$_2$BaNi(PO$_4$)$_2$ has been discussed in several recent works~\cite{ShengJ2025_NiP1,ShengJ2025_NiP2,HuangQ2025},
while several subtle details have not been covered yet.

Firstly, as we discussed in the $T=0$ section, the semiclassical SU(3) approach completely misses the SN phases for realistic parameters ($K=0$) of Na$_2$BaNi(PO$_4$)$_2$.
As a result, calculations of spin dynamics in the SN phases cannot directly use the GLSW approach~\cite{MatveevVM1973,PapanicolaouN1988,TsunetsuguH2006,LauchliA2006,MunizRA2014} which is based on the SU(3) coherent states.
Since we have demonstrated in Fig.~\ref{fig:phd_T0} that a finite yet small biquadratic term helps to stabilize the SN phases, we can compute the spin dynamics of the SN phases with finite $K$, to gain insight into the key qualitative physics of this compound.

\begin{figure}
\centering
\includegraphics[width=\columnwidth]{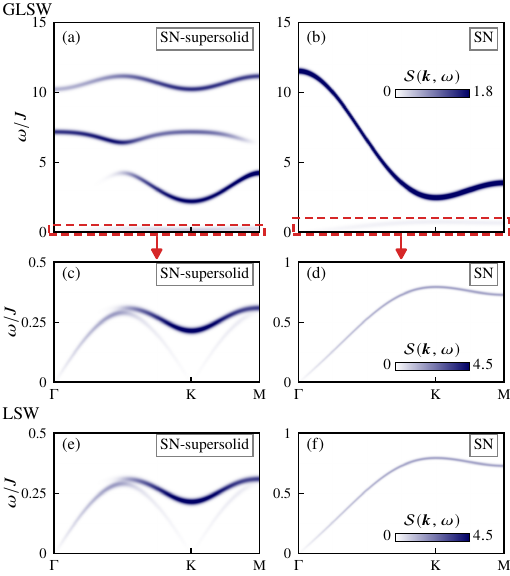}
\caption{(a)(b) GLSW calculation of the DSSF for the spin-1 model Eq.~\eqref{eq:H_tot} in the SN-supersolid ($H / J=0.02$) and SN ($H / J =6.5$) phases, at $\Delta=1$, $D/J=5$ and $K/J=-0.1$.
(c)(d) Zoomed-in view of the low-energy part as highlighted by the dashed rectangles in (a)(b).
(e)(f) LSW calculation of the DSSF for the effective $s=1/2$ model Eq.~\eqref{eq:Heff} in the SN-supersolid ($\tilde{H} / | \tilde{J} | = 0.2$) and SN ($\tilde{H} / | \tilde{J} | = 65$) phases, at $\tilde{\Delta}=-21$.}
\label{fig:Sqw_SN}
\end{figure}

%To map the corresponding excitations, we select representative magnetic fields $H/J = 0.02$, $3$ and $6.5$ traversing the distinct phases.
%The total DSSF $\mathcal{S}(\bm{k}, \omega) =\mathcal{S}^{xx}(\bm{k}, \omega)+\mathcal{S}^{yy}(\bm{k}, \omega) + \mathcal{S}^{zz}(\bm{k}, \omega)$ is calculated via
%\begin{equation}
%\mathcal{S}^{\alpha \beta}\left( \bm{k}, \omega \right) = 2\pi \sum_{\varphi \neq 0} \langle 0 | S_{\bm{k}}^{\alpha} | \varphi \rangle \langle \varphi | S^{\beta}_{- \bm{k}} | 0 \rangle \delta(\omega + \omega_0 - \omega_{\varphi}),
%\end{equation}
%where $| 0 \rangle$ represents the corresponding quasi-particle vacuum and $| \varphi \rangle$ denotes intermediate excited states possessing discrete energies $\omega_{\varphi}$.
%The ideally singular $\delta(\omega)$ response is dynamically evaluated employing a standard Lorentzian broadening function $L(x) = \sigma/ [ \pi ( x^2 + \sigma^2 )]$.
%Crucially, while deploying Gaussian smoothing is more broadly prevalent, it would severely obscure the delicate low-energy dispersion details uniquely critical to analyzing SN phase properties.

Figures~\ref{fig:Sqw_SN}(a)(b) show the calculated dynamic spin structure factor (DSSF) $\mathcal{S}(\bm{k},\omega)\equiv \sum_\alpha S^{\alpha \alpha}(\bm{k},\omega)$ with $\Delta=1$, $D/J=5$, $K/J=-0.1$ and two fields: $H/J=0.02$ (SN-supersolid) and $H/J=6.5$ (SN).
Gapless Goldstone modes (quadrupole waves) are observed in both SN phases.
At low energy [Figs.~\ref{fig:Sqw_SN}(c)(d)] the intensity of the gapless modes vanishes at the ordering wave vector, which agrees with the quadrupole nature of the in-plane components of the SN phases~\cite{ShengJ2025_NiP2}.

In Ref.~\cite{ShengJ2025_NiP2}, the low-energy quadrupole waves were computed via LSW for the effective $s=1/2$ model Eq.~\eqref{eq:Heff} to avoid the instability of the GLSW method, and the results are recomputed in Figs.~\ref{fig:Sqw_SN}(e)(f), with $\tilde{\Delta}=-21$ and two fields: $\tilde{H}/|\tilde{J}|=0.2$ (SN-supersolid) and $\tilde{H}/|\tilde{J}|=65$ (SN).
As emphasized in Ref.~\cite{ShengJ2025_NiP2}, for the effective spin-$1/2$ model, only the longitudinal components of the DSSF contribute to that of the original model:
\begin{equation}
\mathcal{S}(\bm{k}, \omega) \approx 4\mathcal{S}^{zz}_\text{eff}(\bm{k}, \omega).
\end{equation}

It is interesting to see that the GLSW (small finite $K$) and the LSW ($K=0$) approaches agree on the quadrupole wave excitations to a substantial degree, confirming that the biquadratic interaction with $K/J=-0.1$ perturbs the low-energy spectra in a negligible manner.

Another interesting feature is the gapped mode around $\omega/J\sim 0.22$ at the K point in Figs.~\ref{fig:Sqw_SN}(c) and (e), which is actually a pseudo-Goldstone mode~\cite{WeinbergS1972,RauJG2018_pseudo} related to the accidental degeneracy we discussed earlier in this paper.
Indeed, if we further reduce the magnetic field to zero (where the accidental degeneracy becomes exact at the semiclassical level), then this pseudo-Goldstone mode will be pushed down to exact zero energy.
In other words, the accidental degeneracy (additional soft directions beyond the U(1)-symmetry breaking) is indeed reflected in the spin excitation spectrum as additional soft modes.
We should note that such accidental degeneracy is only true at the semiclassical level at $T=0$, and will be lifted by quantum or thermal fluctuations~\cite{VillainJ1977_order_disorder,VillainJ1980_order_disorder,ShenderEF1982,ShengQ1992,SeifertUFP2022}.
In fact, as shown in Figs.~\ref{fig:Sqw_SN}(c) and (e), a magnetic field is found to open a gap in the pseudo-Goldstone mode, and the SN-supersolid state becomes the true ground state.

\section{Discussion}
\label{sec:Discussion}

In summary, we have investigated the finite-temperature behaviors and dynamics of a TL spin-1 model with large single-ion anisotropy, relevant to the compound Na$_2$BaNi(PO$_4$)$_2$ which recently shows strong evidence of SN physics.

Several physical insights were gained in our work.
Firstly, we have shown that the SN phases can be further stabilized by finite biquadratic interactions.
At the semiclassical SU(3) level, now we are able to write down the wave functions of the SN-supersolid and SN phases, which further reveals the accidental degeneracy and its associated mathematical structure at zero field.

The finite-T phase diagram of the Na$_2$BaNi(PO$_4$)$_2$ can be captured by both the semiclassical SU(3) approach for the original model (semiclassical MC) and the effective spin-$1/2$ model (QMC).
In both cases, we see a SN-supersolid phase and a SN phase emerging as low-temperature phases of the model.
In particular, in the experiments for Na$_2$BaNi(PO$_4$)$_2$ the SN-supersolid/UUD boundary has not been identified due to its low temperature, and the phase diagrams established in this paper provide helpful predictions of its shape.

It is also interesting to note that there was no clear specific heat anomaly reported for the first-order UUD/SN phase boundary of Na$_2$BaNi(PO$_4$)$_2$~\cite{ShengJ2025_NiP1}.
In this paper, the UUD/SN boundary is found to be practically a vertical first order line, which explains why there is no anomaly observed in the experimental specific heat yet: perfectly vertical boundary corresponds to anomaly in partition function derivatives with respect to $H$ instead of $T$ so there is no specific heat anomaly at all; while almost-vertical boundary requires very fine magnetic field steps not to miss it.

Our spin dynamics calculation for the original $S=1$ model confirms the LSW results of Ref.~\cite{ShengJ2025_NiP2}, demonstrating the existence of the quadrupole waves.
Moreover, this calculation reveals the connection between the pseudo-Goldstone mode and the accidental degeneracy of the $T=0$ semiclassical states at zero field.

Finally, we should emphasize that the models studied in this paper are two-dimensional, which is why BKT arguments should be applied.
Actually, in the Na$_2$BaNi(PO$_4$)$_2$ compound tiny inter-layer interactions are present, which should replace the BKT transition by a transition into 3D long-range ordered states, and small logarithmic corrections should be considered~\cite{ShengJ2025_NiP1}.

%
%
%The semi-classical numerical simulations in this work were performed utilizing the SpinToolkit package.~\cite{sptk}

\begin{acknowledgments}
We thank Jieming Sheng and Liusuo Wu for valuable discussions and collaborations on this topic.
This work was supported by the National Key Research and Development Program of China (Grants No.~2024YFA1408303 and No.~2025YFA1411503), the National Natural Science Foundation of China (Grants No.~12374124), and the Fundamental Research Funds for the Central Universities (Grant No.~226-2024-00068).
Part of the calculations were performed on Quantum Many-Body HPC platform at School of Physics, Zhejiang University.
\end{acknowledgments}

\bibliography{refs}

%\clearpage
\appendix

\section{Conversion between dipolar and quadrupolar bases}
\label{sec:conversion}

When working with quantum spin $S=1$ states, two conventions are commonly adopted.
One typical choice is to expand the local wave function at site-$i$ using the dipolar basis:
\begin{equation}\label{eq:Z_convention}
|\Psi_i\rangle = \{ |1\rangle,\,  |0\rangle,\, |-1\rangle \} \bm{Z}_i,
\end{equation}
where $|1\rangle$, $|0\rangle$, and $|-1\rangle$ are the eigenstates of the $S_i^z$ operator, and the vector $\bm{Z}_i$ has three complex components:
\begin{equation}
\bm{Z}_i =
\begin{pmatrix}
Z_i^0 \\
Z_i^1 \\
Z_i^2
\end{pmatrix}.
\end{equation}

Another choice is widely adopted in the study of quadrupolar orders, using a time-reversal-symmetric quadrupolar basis~\cite{IvanovBA2003,LauchliA2006,SmeraldA2013}:
\begin{equation}\label{eq:d_convention}
|\Psi_i\rangle = \{ |x\rangle,\,  |y\rangle,\, |z\rangle \} \bm{d}_i,
\end{equation}
where
\begin{equation}
| x\rangle = \iu\frac{|1\rangle-|-1\rangle}{\sqrt{2}},\,| y\rangle=\frac{|1\rangle+|-1\rangle}{\sqrt{2}},\,| z\rangle=-\iu|0\rangle,
\end{equation}
and the vector $\bm{d}_i$ (director) is denoted by three complex components:
\begin{equation}\label{eq:d_components}
\bm{d}_i =
\begin{pmatrix}
d_i^x \\
d_i^y \\
d_i^z
\end{pmatrix}.
\end{equation}

In both cases, six real parameters are involved.
After imposing the normalization condition and fixing the overall phase of the wave function, only four of them are independent.
While both conventions are legitimate, and using either one should lead to the same physical results, a few differences remain:
\begin{enumerate}
\item There is a unitary transformation between the two bases;
\item There is a freedom in fixing the global phase in each convention.
\end{enumerate}

In this work, the semiclassical analyses with SU(3) coherent states are carried out with the \textit{SpinToolkit} library whose engine builds upon the dipolar basis.
On the other hand, the quadrupolar orders are the key object of interest in this paper.
So here we provide explicitly the conversion from $\bm{Z}_i$ to $\bm{d}_i$ for convenience.

As a first step, by comparing Eqs.~\eqref{eq:Z_convention} and \eqref{eq:d_convention}, we can obtain a ``raw'' director $\bm{d}_i^{\text{raw}}$ from a given $\bm{Z}_i$:
\begin{equation}
\bm{d}_i^{\text{raw}}
= U \bm{Z}_i,
\end{equation}
where
\begin{equation}
U =
\begin{pmatrix}
-\iu/\sqrt{2} & 0 & \iu/\sqrt{2}\\
1/\sqrt{2} & 0 & 1/\sqrt{2}\\
0 & \iu & 0
\end{pmatrix}.
\end{equation}

As the transformation matrix $U$ is unitary, the normalization condition $|\bm{d}_i^{\text{raw}}|^{2}=|\bm{Z}_i|^{2}=1$ is already satisfied.
While $\bm{d}_i^{\text{raw}}$ represents the same quantum state, its overall phase has not been fixed to follow the commonly adopted gauge choice of the director.

\subsection{Gauge choice 1: request \texorpdfstring{$\bm{d}_i \cdot \bm{d}_i$}{d_i dot d_i} to be real}
In the literature for studying the quadrupolar orders~\cite{IvanovBA2003,LauchliA2006,SmeraldA2013}, a common convention is to separate the director $\bm{d}_i$ into real and imaginary parts,
\begin{equation}
\bm{d}_i=\bm{u}_i+\iu \bm{v}_i,
\end{equation}
and fix the gauge by requesting
\begin{equation}
\bm{u}_i \cdot \bm{v}_i=0.
\label{eq:uDOTv}
\end{equation}

Since the scalar product
\begin{equation}
\bm{d}_i \cdot \bm{d}_i = |\bm{u}_i|^2 - |\bm{v}_i|^2 + 2 \iu \bm{u}_i \cdot \bm{v}_i,
\end{equation}
it is clear that the condition Eq.~\eqref{eq:uDOTv} is equivalent to
\begin{equation}
\text{Im}(\bm{d}_i \cdot \bm{d}_i) = 0.
\end{equation}

As a result, we can fulfill Eq.~\eqref{eq:uDOTv} by performing a U(1) rotation of $\bm{d}_i^{\text{raw}}$:
\begin{equation}
\bm{d}_i = e^{\iu \varphi} \bm{d}_i^{\text{raw}},
\end{equation}
where
\begin{equation}
\varphi = -\frac{1}{2} \arg (\bm{d}_i^{\text{raw}} \cdot \bm{d}_i^{\text{raw}}).
\end{equation}

\subsection{Gauge choice 2: request \texorpdfstring{$d_i^z$}{d_i^z} to be real}

Alternatively, we can fix the overall phase by imposing~\cite{SeifertUFP2022}
\begin{equation}
\text{Im}(d_i^z) = 0.
\end{equation}

This can be achieved by a U(1) rotation to gauge away the imaginary part of $(d_i^{z})^\text{raw}$:
\begin{equation}
\bm{d}_i = e^{\iu \varphi} \bm{d}_i^{\text{raw}},
\end{equation}
where
\begin{equation}
\varphi = -\arg (d_i^z)^\text{raw}.
\end{equation}

\section{Degeneracy of the zero-field phase}
\label{sec:degeneracy}

In this section, we consider the spin-$1$ model at $H=0$, where the semiclassical SU(3) analysis shows a large degeneracy for strongly easy-axis single-ion anisotropy $D/J\gg 1$.
This degeneracy was reported by Seifert and Savary~\cite{SeifertUFP2022}, where they named the family of degenerate states as zAFM$_{\mathrm{deg}}$, and argued that supersolidity can be realized in the presence of an infinitesimal magnetic field.
Indeed, our calculation under a finite magnetic field does support this mechanism, where the SN-supersolid occupies a finite region in Fig.~\ref{fig:phd_T0}(b) of the main text.

The purpose of this section is to show the mathematical structure of the degeneracy explicitly, following the trick by Kleine \textit{et al.}~\cite{KleineB1992} which analyzed the degeneracy of the SU(2) semiclassical states of the antiferromagnetic XXZ model on TL~\cite{MiyashitaS1985}.

First, recall that the directors $\bm{d}_s$ that minimize the SN phases are already parameterized in Eq.~\eqref{eq:director-SN} of the main text.
By setting the spontaneously broken U(1) angle $\phi_s=0$, and introducing another variable $\eta_s \equiv \chi_s + \pi/4$, we obtain
\begin{equation}
\bm{d}_s =
\begin{pmatrix}
\cos \left( \eta_s - \frac{\pi}{4} \right) \\
\iu \cos \left( \eta_s + \frac{\pi}{4} \right) \\
0
\end{pmatrix},
\end{equation}
which recovers the parameterization of the zAFM$_{\mathrm{deg}}$ state in Ref.~\cite{SeifertUFP2022}.

A straightforward evaluation shows that the energy density of the zAFM$_{\mathrm{deg}}$ state at $H=0$ is:
\begin{align}
e&=
\left(\Delta J-\frac{K}{2}\right) (\cos2\eta_{A}\cos2\eta_{B}+\cos2\eta_{B}\cos2\eta_{C}  \nonumber \\
&\quad \qquad \qquad +\cos2\eta_{C}\cos2\eta_{A} ) \nonumber \\
&\quad
+\frac{K}{2}\left(\sin2\eta_{A}\sin2\eta_{B}+\sin2\eta_{B}\sin2\eta_{C} + \sin2\eta_{C}\sin2\eta_{A}\right) \nonumber \\
&\quad +\frac{9}{2}K-D.
\end{align}

We further parameterize
\begin{subequations}
\begin{align}
A &\equiv \Delta J - \frac{K}{2},\\
B &\equiv \frac{K}{2},\\
e_0 &\equiv \frac{9}{2}K-D,\\
x_i &\equiv 2\eta_i,
\end{align}
\end{subequations}
then the energy density becomes
\begin{equation}
e=\sum_{i<j} \left( A \cos x_i \cos x_j + B \sin x_i \sin x_j \right)  + e_0,
\end{equation}
which is practically the same form as in Kleine \textit{et al.}~\cite{KleineB1992}, suggesting the accidental degeneracy of the zAFM$_{\mathrm{deg}}$ state.

There is one subtle difference though.
Define the single dimensionless ratio
\begin{equation}
\lambda \equiv \frac{B}{A}.
\end{equation}

In the paper by Kleine \textit{et al.}, $\lambda$ is simply the exchange anisotropy $\Delta$ which is positive.
In our current case we have $\lambda = K/(2\Delta J - K) < 0$ for $K<0$, corresponding to the ferromagnetic transverse exchange that has been analyzed by Murthy \textit{et al}~\cite{MurthyG1997}.
Following Kleine and Murthy, we rewrite
\begin{equation}
e = \frac{A}{2}(1+\lambda) \sum_{i<j} \cos(x_i - x_j)  + \frac{A}{2}(1-\lambda) \sum_k \cos \left(x_k - X  \right) + e_0,
\end{equation}
where
\begin{equation}
X\equiv x_A + x_B + x_C.
\end{equation}

Finally, by introducing the planar unit vectors
\begin{equation}
\bm{n}_i \equiv
\begin{pmatrix}
\sin \left( x_i - X \right) \\
\cos \left( x_i - X \right)
\end{pmatrix},
\end{equation}
and the sum
\begin{equation}
\bm{m} \equiv \bm{n}_A + \bm{n}_B + \bm{n}_C,
\end{equation}
we obtain
\begin{equation}\label{eq:energy_degeneracy}
e = A \left[ \frac{1+\lambda}{4} \left(|\bm{m}|^2 - 3 \right) + \frac{1-\lambda}{2}m_y \right] + e_0,
\end{equation}
which is essentially Kleine and Murthy's results generalized to the $S=1$ model.

The accidental degeneracy becomes clear by counting the number of degrees of freedom.
Equation~\eqref{eq:energy_degeneracy} depends on three variables $\eta_A$, $\eta_B$, $\eta_C$ only through two numbers $m_x$ and $m_y$.
Therefore, the energy minimum is generically a one-parameter manifold that parameterizes the accidental degeneracy.

\onecolumngrid

\section{Spin stiffness}
\label{sec:stiffness}

In this section, we derive the spin stiffness expression in the semiclassical approach.

With identity
\begin{equation}
\left( \bm{S}_i \cdot \bm{S}_j\right)^2 = \frac{4}{3} - \frac{1}{2} \bm{S}_i \cdot \bm{S}_j + \frac{1}{2} \bm{Q}_i \cdot \bm{Q}_j
\end{equation}
for $S=1$, we can rewrite
\begin{align}
\mathcal{H}_\text{tot} &= \left( J - \frac{K}{2}\right) \sum_{\langle ij \rangle} \left(S_i^x S_j^x + S_i^y S_j^y \right) + \left( \Delta J - \frac{K}{2}\right) \sum_{\langle ij \rangle} S_i^z S_j^z
+ \frac{K}{2} \sum_{\langle ij \rangle} \bm{Q}_i \cdot \bm{Q}_j \nonumber \\
&\quad -D\sum_i (S_i^z)^2 -H \sum_i S_i^z + 4NK \nonumber \\
&= \left( \frac{J}{2} - \frac{K}{4}\right) \sum_{\langle ij \rangle} \left(S_i^+ S_j^- + S_i^- S_j^+ \right) + \left( \Delta J - \frac{K}{2}\right) \sum_{\langle ij \rangle} S_i^z S_j^z \nonumber \\
&\quad + \frac{K}{4} \sum_{\langle ij \rangle} \left( Q_i^{+2}Q_j^{-2} + Q_i^{-2}Q_j^{+2} + Q_i^{+1}Q_j^{-1} + Q_i^{-1}Q_j^{+1}  \right) + \frac{K}{2} \sum_{\langle ij \rangle}  Q_i^{3z^2-r^2} Q_j^{3z^2-r^2} \nonumber \\
&\quad -D\sum_i (S_i^z)^2 -H \sum_i S_i^z + 4NK.
\end{align}

For a twist
\begin{equation}
|\Psi_i\rangle \rightarrow e^{-\iu \theta_i S_i^z} |\Psi_i\rangle,\quad \theta_i \equiv \varphi (\hat{\bm{e}} \cdot \bm{r}_i),
\end{equation}
the semiclassical energy per bond $e_{ij}$ becomes:
\begin{equation}\label{eq:e_per_bond}
e_{ij} = \left( J - \frac{K}{2}\right) \text{Re} \left[ e^{\iu \theta_{ij}} \langle S_i^+ \rangle \langle S_j^- \rangle \right]
+ \frac{K}{2} \text{Re} \left[ e^{2\iu \theta_{ij}} \langle Q_i^{+2} \rangle \langle Q_j^{-2} \rangle + e^{\iu \theta_{ij}} \langle Q_i^{+1} \rangle \langle Q_j^{-1} \rangle \right] + c_{ij},
\end{equation}
where the twist-invariant part
\begin{equation}
c_{ij} = \left( \Delta J - \frac{K}{2}\right)  \langle S_i^z \rangle \langle S_j^z \rangle + \frac{K}{2} \langle Q_i^{3z^2-r^2} \rangle \langle Q_j^{3z^2-r^2} \rangle + \frac{4}{3}K,
\end{equation}
and
\begin{equation}
\theta_{ij} \equiv \theta_i - \theta_j.
\end{equation}

Differentiating at $\theta_{ij}=0$, we obtain a bond current term $j_{ij}$ and a bond kinetic energy term $T_{ij}$:
\begin{subequations}
\begin{align}
j_{ij} &\equiv \left. \frac{\partial e_{ij}}{\partial \theta_{ij}} \right|_{\theta_{ij}=0} = - \left( J - \frac{K}{2}\right) \text{Im} \left[  \langle S_i^+ \rangle \langle S_j^- \rangle \right]  - K \text{Im} \left[  \langle Q_i^{+2} \rangle \langle Q_j^{-2} \rangle \right] - \frac{K}{2} \text{Im} \left[  \langle Q_i^{+1} \rangle \langle Q_j^{-1} \rangle \right], \\
T_{ij} &\equiv -\left. \frac{\partial^2 e_{ij}}{\partial \theta_{ij}^2} \right|_{\theta_{ij}=0} =  \left( J - \frac{K}{2}\right) \text{Re} \left[  \langle S_i^+ \rangle \langle S_j^- \rangle \right]  + 2 K \text{Re} \left[  \langle Q_i^{+2} \rangle \langle Q_j^{-2} \rangle \right] + \frac{K}{2} \text{Re} \left[  \langle Q_i^{+1} \rangle \langle Q_j^{-1} \rangle \right].
\end{align}
\end{subequations}

The free energy is
\begin{equation}
F=-T \ln \int \prod_i \mathrm{d}\mu (\Psi_i) e^{-\beta E},
\end{equation}
where $\mathrm{d}\mu(\Psi_i)$ is the measure and  $E$ is the semiclassical internal energy of the twisted configuration.

The derivatives of the free energy are:
\begin{subequations}
\begin{align}
\frac{\partial F}{\partial \varphi} &= \left\langle \frac{\partial E}{\partial \varphi} \right\rangle_T, \\
\frac{\partial^2 F}{\partial \varphi^2} &= \left\langle \frac{\partial^2 E}{\partial \varphi^2} \right\rangle_T - \beta \left[ \left\langle \left( \frac{\partial E}{\partial \varphi} \right)^2 \right\rangle_T - \left\langle \frac{\partial E}{\partial \varphi} \right\rangle_T^2 \right].
\end{align}
\end{subequations}
where $\langle \cdots \rangle_T$ denotes the thermal average.

The spin stiffness is then obtained via
\begin{align}
\rho_s &= \frac{1}{\mathcal{A}} \left. \frac{\partial^2 F}{\partial \varphi^2} \right|_{\varphi=0} \nonumber \\
&= \frac{2}{\sqrt{3}N} \left\{ - \left\langle \sum_{\langle ij\rangle}   \left( \hat{\bm{e}} \cdot \bm{r}_{ij} \right)^2 T_{ij} \right\rangle_T -\beta \left[  \left\langle \mathcal{J}_{\hat{\bm{e}}}^2 \right\rangle_T - \left\langle \mathcal{J}_{\hat{\bm{e}}} \right\rangle_T^2 \right] \right\},\label{eq:stiffness}
\end{align}
where $\mathcal{A}=\sqrt{3}N/2$ is the total area of the TL, and the total current
\begin{equation}
\mathcal{J}_{\hat{\bm{e}}} \equiv \left. \frac{\partial E}{\partial \varphi} \right|_{\varphi=0} = -\sum_{\langle ij \rangle}  \left( \hat{\bm{e}} \cdot \bm{r}_{ij} \right) j_{ij},
\end{equation}
and
\begin{equation}
\bm{r}_{ij} \equiv \bm{r}_j - \bm{r}_i.
\end{equation}

At this point, we note that Eq.~\eqref{eq:stiffness} applies to the general Hamiltonian $\mathcal{H}_\text{tot}$, where the biquadratic term is present.

\subsection{$K=0$ case}
Consider the $K=0$ limit, in which case we can simplify
\begin{subequations}
\begin{align}
j_{ij} &\stackrel{K=0}{=} -  J  \text{Im} \left[  \langle S_i^+ \rangle \langle S_j^- \rangle \right]
        = J \left( \langle \bm{S}_i \rangle \times \langle \bm{S}_j \rangle \right)^z, \\
T_{ij} &\stackrel{K=0}{=}   J  \text{Re} \left[  \langle S_i^+ \rangle \langle S_j^- \rangle \right]
        = J \left( \langle S_i^x \rangle \langle S_j^x \rangle + \langle S_i^y \rangle \langle S_j^y \rangle \right).
\end{align}
\end{subequations}

Consequently, the spin stiffness is reduced to:
\begin{align}
\rho_s &\stackrel{K=0}{=} -\frac{2J}{\sqrt{3}N} \left\langle \sum_{\langle ij\rangle} \left( \hat{\bm{e}} \cdot \bm{r}_{ij}\right)^2 \left( \langle S_i^x \rangle \langle S_j^x \rangle + \langle S_i^y \rangle \langle S_j^y \rangle \right) \right\rangle_T \nonumber \\
&\quad - \frac{2\beta J^2}{\sqrt{3}N} \left\{
\left\langle \left[ \sum_{\langle ij\rangle} \left( \hat{\bm{e}} \cdot \bm{r}_{ij}\right) \left( \langle \bm{S}_i \rangle \times \langle \bm{S}_j \rangle \right)^z \right]^2 \right\rangle_T
-\left\langle \sum_{\langle ij\rangle} \left( \hat{\bm{e}} \cdot \bm{r}_{ij}\right) \left( \langle \bm{S}_i \rangle \times \langle \bm{S}_j \rangle \right)^z \right\rangle_T^2
\right\}.
\end{align}

\section{Free energy cost of a twist}
\label{sec:twist}

We give the $2L$ seam bonds of Fig.~\ref{fig:stiffness}(b) in the main text a phase $\alpha$ in the charge-1 channel only, with every seam bond $(ij)$ oriented in the same direction across the cut.
According to Eq.~\eqref{eq:e_per_bond}, the charge-1 part of the bond energy is:
\begin{equation}
a_{ij} \equiv \left( J - \frac{K}{2}\right) \langle S_i^+ \rangle \langle S_j^- \rangle + \frac{K}{2} \langle Q_i^{+1} \rangle \langle Q_j^{-1} \rangle ,
\end{equation}
and the seam replaces it by $\text{Re} \left[ e^{\iu\alpha} a_{ij} \right]$, leaving all other terms, including the charge-2 channel, unchanged.
The energy of a configuration and its derivative are therefore
\begin{subequations}
\begin{align}
E(\alpha) &= E(0) + \sum_{(ij)\in\text{seam}} \text{Re} \left[ \left( e^{\iu\alpha} - 1 \right) a_{ij} \right], \\
\frac{\partial E}{\partial \alpha} &= -\sum_{(ij)\in\text{seam}} \text{Im} \left[ e^{\iu\alpha} a_{ij} \right].
\end{align}
\end{subequations}
Since the measure $\prod_i \mathrm{d}\mu(\Psi_i)$ does not depend on $\alpha$, the derivation of $\partial F/\partial\alpha$ in Sec.~\ref{sec:stiffness} carries over, and
\begin{equation}
\delta F(\alpha) \equiv F(\alpha) - F(0) = \int_0^{\alpha} \left\langle \frac{\partial E}{\partial \alpha'} \right\rangle_T \mathrm{d}\alpha' ,
\end{equation}
where $\langle \cdots \rangle_T$ is the thermal average with seam angle $\alpha'$.

At $\Delta=1$, $D/J=5$, $K/J=-0.2$, $H/J=6.75$, and $T/J=0.008$, we computed $\langle \partial E/\partial\alpha' \rangle_T$ at 21 equally spaced angles $\alpha'_n = n\pi/20$ ($n=0,\dots,20$), each in an independent simulation with $10^5$ thermalization sweeps and $10^6$ measurement sweeps, measuring every 20 sweeps.
The error $\sigma_n$ of each mean comes from a blocking analysis with blocks of at least 20 integrated autocorrelation times.
The symbols in Fig.~2(d) are the cumulative Simpson integrals $\delta F(\alpha_n) = \sum_m \left. w_{nm}  \partial E/\partial\alpha' \right|_{\alpha'_m}$, where $w_{nm}$ are the quadrature weights; because the simulations are independent, $\sigma^2[\delta F(\alpha_n)] = \sum_m w_{nm}^2 \sigma_m^2$.
The endpoint is consistent with zero at every size: $\delta F(\pi)/J = -4.7(4.3)$, $+3.3(5.2)$, $-0.8(6.0)$, $-6.3(8.3)$, and $+8.0(10.3)$, in units of $10^{-5}$, for $L=24$, 36, 48, 72, and 96.
The empty circles in Fig.~2(e) therefore show upper limits,
\begin{equation}
\delta F_\text{max} = \max \left\{ \left| \delta F(\pi) \right| + 3\sigma\left[ \delta F(\pi) \right] + \left| \delta F_\text{S}(\pi) - \delta F_\text{T}(\pi) \right| ,\ \delta F_\text{sys} \right\},
\end{equation}
where $\delta F_\text{S}$ and $\delta F_\text{T}$ are the Simpson and trapezoidal integrals of the same 21 points, whose difference measures the discretization error.
The floor $\delta F_\text{sys} = 3.4\times10^{-4} J$ is an empirical systematic error, taken from the same calculation above the BKT transition at $T/J=0.016$: there $\delta F(\pi)$ should vanish as $L\to\infty$, but we found $-0.8(9)\times10^{-4} J$ at $L=24$ and $-3.4(1.2)\times10^{-4} J$ at $L=48$.
This gives $\delta F_\text{max} = 3.4\times10^{-4} J$ at $L=24$, 48, and 72, where the floor applies, and $4.4\times10^{-4} J$ at $L=96$.
These limits lie at least a factor of 100 below the cheaper of the two ordered-sector costs.

In Fig.~\ref{fig:stiffness}(e) of the main text, we estimated the Ising wall tension per bond via
\begin{equation}
\sigma_\text{I}^\text{est} \equiv \frac{1}{2L} \langle E(\pi) - E(0) \rangle_T  = -2 \langle \text{Re}\, a_{ij} \rangle_T,
\end{equation}
i.e., the mean energy per bond of reversing the charge-1 coupling evaluated at temperature $T$ in the $\alpha=0$ ensemble.
A separate simulation at $L=6$ gives $\beta\sigma_\text{I}^\text{est} = 0.114$ at $T/J=0.02$, rising to $0.126$ at $T/J=0.005$; we use the smaller value, $\sigma_\text{I}^\text{est} = 0.114\,T$, as the scale of the wall cost.
The dotted line is $\pi^2\rho_s/\sqrt{3} = 0.33 J$, with $\rho_s$ taken from Fig.~2(a) at $T/J=0.008$.

\section{Finite-temperature phase boundaries}
\label{sec:boundaries}

Figure~\ref{fig:phd_SU3} of the main text contains three kinds of phase boundary, which we locate by different methods: the three-state Potts transition of the $S^z$ sector, the BKT transition of the quadrupolar sector, and the first-order boundary between the UUD and SN states.
All three use the semiclassical MC simulations described in the main text, at $\Delta=1$, $D/J=5$, and $K/J=-0.2$, with $L=24$, 36, 42, and 48 unless stated otherwise.
Figure~\ref{fig:boundaries} illustrates each method at one representative parameter set.

\subsection{Three-state Potts boundary}

The Potts boundary is located with the Binder parameter $U_{\mathcal{S}_\text{d}^{zz}(\text{K})}$ defined in the main text.
The associated critical temperature $T_{c2}$ is determined from a finite-size-scaling collapse of different system sizes,
\begin{equation}\label{eq:binder_collapse}
U_{\mathcal{S}_\text{d}^{zz}(\text{K})} = f\left[ (T-T_{c2}) L^{1/\nu} \right],
\end{equation}
with $\nu=5/6$ fixed at its exact value for the two-dimensional three-state Potts universality class~\cite{WuFY1982_RMP}.
We approximate $f$ by a polynomial of degree 3, 4, or 5 and, for each trial $T_{c2}$, fit it by the weighted least-squares method.

Figure~\ref{fig:boundaries}(a) shows the finite-size scaling at $H/J=3$, where $T_{c2}/J = 0.5243(7)$ leads to a good collapse of the data at different $L$.

\subsection{BKT boundary}

%The BKT boundary is located with the stiffness $\rho_s$ of Sec.~\ref{sec:stiffness}.
%For each $L$ we find the temperature $T^*(L)$ at which $\rho_s(T)$ crosses $8T/\pi$, the universal value for the unbinding of half vortices [Fig.~2(a) of the main text], by linear interpolation between neighboring temperatures.
%Its uncertainty is the larger of half the local temperature spacing and the half-width of the range of crossings obtained from $\rho_s \pm \sigma_{\rho_s}$.
Near a BKT transition the correlation length diverges as $\xi \sim \exp(b\,t^{-1/2})$ with $t=T/T_{c1}-1$~\cite{KosterlitzJM1974}, and setting $\xi[T^*(L)] \sim L$ gives the logarithmic finite-size form~\cite{HasenbuschM2005}
\begin{equation}
T^*(L) = T_{c1} \left[ 1 + \frac{b^2}{\ln^2(L/L_0)} \right].
\end{equation}
With four sizes ($L=\{24,36,42,48\}$) spanning only a factor of two, $b^2$ and $L_0$ cannot be determined separately: $\chi^2$ is essentially unchanged for $0.1 \le b^2 \le 8$.
We therefore fix $b=1$ and fit $T_{c1}$ and $L_0$, weighting each $T^*(L)$ by its uncertainty.
%The error of $T_{c1}$ combines in quadrature the standard deviation over 2000 parametric bootstrap resamples of $T^*(L)$ and the difference from an alternative fit with $L_0=1$ and $b^2$ free.

\begin{figure}[tbp]
\centering
\includegraphics[width=\textwidth]{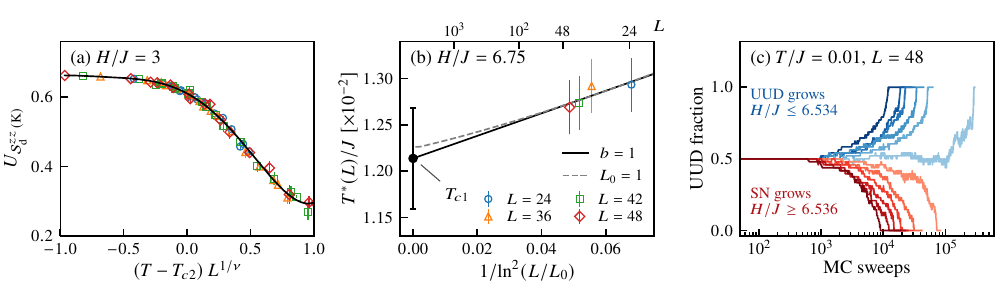}
\caption{(a) Binder parameter collapse as in Eq.~\eqref{eq:binder_collapse} at $H/J=3$.
The solid line is the degree-5 polynomial fit.
(b) Crossings $T^*(L)$ of $\rho_s$ with $8T/\pi$ against $1/\ln^2(L/L_0)$ at $H/J=6.75$, with $L$ on the top axis.
The solid line is the fit $T^*(L) = T_{c1}[1+1/\ln^2(L/L_0)]$, and the filled circle is its extrapolation $T_{c1}$ with error; the dashed line is the alternative fit with $L_0=1$ and $b^2$ free.
(c) Evolution of states prepared from half-UUD and half-SN at $T/J=0.01$ and $L=48$. Twelve fields spaced by $0.002J$ are shown, where darker colors are farther from $H_{c2}$.}
\label{fig:boundaries}
\end{figure}

At $H/J=6.75$ [Fig.~\ref{fig:boundaries}(b)], $T^*(L)/J$ decreases from $0.01293$ at $L=24$ to $0.01269$ at $L=48$ and extrapolates to $T_{c1}/J = 0.0121(5)$, with $L_0=0.52$.
%Over the 12 fields in Fig.~3, the extrapolated $T_{c1}$ lies 2--7\% below $T^*(48)$.

\subsection{First-order boundary}

The transition between the UUD and SN states is first order: at $T=0$ the energies of the two states cross at $H_{c2}/J\approx 6.557$, where the magnetization jumps.

At $T>0$ we locate it by direct coexistence.
Each simulation starts from a configuration in which half of the rows of the $L\times L$ lattice are in the $T=0$ UUD state and the other half in the uniform SN state, and runs at fixed $T$ and $H$.
The phase with the lower free energy grows until it fills the system.
For each temperature we scan twelve fields spaced by $0.002J$, with two independent runs per field at $L=24$, 36, and 48.
We quote $H_{c2}$ as the midpoint between the highest field at which every run ends in the UUD state and the lowest field at which every run ends in the SN state, with half that interval as the error.
This gives $H_{c2}/J = 6.546(2)$, 6.535(1), 6.522(2), and 6.510(2) at $T/J=0.005$, 0.01, 0.015, and 0.02, and the three sizes give consistent brackets [see Fig.~\ref{fig:boundaries}(c)].

\section{Effective Model}
\label{sec:effective_model}

In the main text, we used the effective spin-$1/2$ model to corroborate the QMC calculation and low-energy spin dynamics.
This section utilizes the Schrieffer-Wolff transformation~\cite{SchriefferJR1966,BravyiS2011} to derive the low-energy effective spin-$1/2$ model from the original spin-$1$ model $\mathcal{H}_\text{tot}$ of the main text.
Setting $K=0$ recovers the model used for the QMC and LSW calculations in the main text.

We consider the large-$D/J$ (strongly easy-axis) limit, where the lowest-energy states consist of a doublet $|S_i^z = \pm 1\rangle$.
In this limit, we can separate the original Hamiltonian into two parts:
\begin{equation}
\mathcal{H}_\text{tot} = \mathcal{H}_0 + \mathcal{V},
\end{equation}
where
\begin{subequations}
\begin{align}
\mathcal{H}_0 & = -D\sum_{i}\left(S_{i}^{z}\right)^{2},\\
\mathcal{V} & = \sum_{\langle ij\rangle} \left\{ J\left[ \frac{1}{2} \left({S_{i}^{+}S_{j}^{-}+S_{i}^{-}S_{j}^{+}}\right) + \Delta S_{i}^{z}S_{j}^{z} \right] + K \left(\bm{S}_i \cdot \bm{S}_j \right)^2 \right\}  -H\sum_{i}S_{i}^{z}.
\end{align}
\end{subequations}
Note that in general $\mathcal{V}$ contains both a block-diagonal part $\mathcal{V}_\text{d}$ and a block-off-diagonal part $\mathcal{V}_\text{od}$.

The Schrieffer-Wolff transformation rotates the original Hamiltonian and projects to the low-energy subspace:
\begin{align}
\mathcal{H}_{\text{eff}} & = P_{L}e^{\mathcal{S}} \mathcal{H} e^{-\mathcal{S}}P_{L} \nonumber \\
&= P_{L}\left\{ \mathcal{H} +\left[\mathcal{S},\mathcal{H}\right] +\frac{1}{2}\left[\mathcal{S},\left[\mathcal{S},\mathcal{H}\right]\right] + \cdots \right\} P_{L},
\end{align}
where $P_L$ is the projector onto the low-energy subspace.

Requesting the cancellation of off-diagonal operators at first order,
\begin{equation}
\mathcal{V}_{\text{od}}+\left[\mathcal{S},\mathcal{H}_{0}\right]=0,
\end{equation}
leads to the effective Hamiltonian at second order:
\begin{equation}
\mathcal{H}_{\text{eff}} = P_{L}\mathcal{H} P_{L} + \frac{1}{2}P_{L} [\mathcal{S},  \mathcal{V}] P_{L}.
\end{equation}

More specifically, the matrix elements of the effective Hamiltonian are:
\begin{equation}
\langle L_{\alpha}|\mathcal{H}_{\textrm{eff}}| L_{\beta}\rangle
= \epsilon_{\alpha}^{(L)} \delta_{\alpha\beta} + \langle L_{\alpha} | \mathcal{V} | L_{\beta} \rangle + \frac{1}{2} \sum_{\gamma} \left[ \frac{\langle L_{\alpha} | \mathcal{V} | H_{\gamma} \rangle \langle H_{\gamma} | \mathcal{V} | L_{\beta} \rangle}{\epsilon_{\alpha}^{(L)} - \epsilon_{\gamma}^{(H)}} + \frac{\langle L_{\alpha} | \mathcal{V} | H_{\gamma} \rangle \langle H_{\gamma} | \mathcal{V} | L_{\beta} \rangle}{\epsilon_{\beta}^{(L)} -\epsilon_{\gamma}^{(H)} } \right],
\label{eq:Heff_2nd_order}
\end{equation}
where $| L_{\alpha} \rangle$ and $| H_{\alpha} \rangle$ denote the eigenstates of $\mathcal{H}_0$ in the low- and high-energy subspaces, respectively, with corresponding eigenvalues $\epsilon_{\alpha}^{(L)}$ and $\epsilon_{\alpha}^{(H)}$.

Consider a two-site problem.
The low-energy manifold consists of four states:
\begin{enumerate}
\item $|L_1\rangle = |S_1^z=1, S_2^z=1\rangle$, with a shorthand notation $|L_1\rangle = |11\rangle$;
\item $|L_2\rangle = |1\bar{1}\rangle$ ($\bar{1} \equiv -1$);
\item $|L_3\rangle = |\bar{1}1\rangle$;
\item $|L_4\rangle = |\bar{1}\bar{1}\rangle$.
\end{enumerate}
The unperturbed energies are identical for all the four low-energy states:
\begin{equation}
\epsilon_{\alpha}^{(L)} \equiv \langle L_\alpha | \mathcal{H}_0 | L_\alpha \rangle = -2D.
\end{equation}

The high-energy states $|H_\alpha\rangle$ are separated into two groups:
\begin{enumerate}
\item $|10\rangle$, $|01\rangle$, $|\bar{1}0\rangle$, $|0\bar{1}\rangle$, with energy $\epsilon_\alpha^{(H)}=-D$;
\item $|00\rangle$, with energy $\epsilon_\alpha^{(H)} = 0$.
\end{enumerate}

According to Eq.~\eqref{eq:Heff_2nd_order}, the matrix elements of the effective Hamiltonian are evaluated to:
\begin{equation}
\mathcal{H}_\text{eff} = -2D + \frac{3K}{2} - \frac{(J-K)^2}{4D} +
\begin{pmatrix}
a - 2H & 0 & 0 & 0 \\
0 &  -a & b & 0\\
0 & b &  -a & 0\\
0 & 0 & 0 &  a +2H
\end{pmatrix},
\end{equation}
where
\begin{subequations}
\begin{align}
a &\equiv \Delta J - \frac{K}{2} + \frac{(J-K)^2}{4D},\\
b &\equiv K - \frac{(J-K)^2}{2D}.
\end{align}
\end{subequations}

For an effective spin-$\frac{1}{2}$ Hamiltonian with two sites, the matrix elements in the same low-energy space are:
\begin{align}
\mathcal{H}_{\text{eff}}	&= C_0 + \tilde{J} \sum_{\langle ij\rangle}   \left(\frac{s_{i}^{+}s_{j}^{-}+s_{i}^{-}s_{j}^{+}}{2}+\tilde{\Delta}s_{i}^{z}s_{j}^{z}\right)  -\tilde{H}\sum_{i}s_{i}^{z}  \nonumber \\
&= C_0 +
\begin{pmatrix}
\frac{\tilde{\Delta} \tilde{J}}{4} - \tilde{H} & 0 & 0 & 0\\
0 & -\frac{\tilde{\Delta} \tilde{J}}{4} & \frac{\tilde{J}}{2} & 0\\
0 & \frac{\tilde{J}}{2} & -\frac{\tilde{\Delta} \tilde{J}}{4} & 0\\
0 & 0 & 0 & \frac{\tilde{\Delta} \tilde{J}}{4} + \tilde{H}
\end{pmatrix}.
\end{align}

Comparing the matrix elements, we obtain
\begin{subequations}
\begin{align}
\tilde{J} &= 2K -\frac{(J-K)^2}{D}, \\
\tilde{\Delta} &=-1 - \frac{4\Delta J D}{(J-K)^2-2KD},\\
\tilde{H} &= 2H, \\
C_0 &= -2D + \frac{3K}{2} - \frac{(J-K)^2}{4D}.
\end{align}
\end{subequations}
By setting $K=0$, we recover Eq.~\eqref{eq:eff_param} in the main text.

\end{document}